\documentclass[ floatfix,secnumarabic, amssymb, nobibnotes, fleqn,10pt]{wlscirep}

\usepackage{mathrsfs} 

\usepackage{amssymb, mathtools, amsmath}

\usepackage{graphicx} 
\usepackage{tikz}
\usetikzlibrary{arrows, shapes, positioning, calc}
\usepackage{quantikz}

\usepackage{quantikz}

\usepackage[utf8]{inputenc}
\usepackage[T1]{fontenc}

\title{ Optimization by VarQITE on Adaptive Variational Quantum Kolmogorov-Arnold Network }

\author[1,*]{ Hikaru Wakaura }

\author[2, 3,/ ]{ Rahmat Mulyawan } 
   
\author[ 2, 3, 4,+ ]{ Andriyan B. Suksmono }  
   
 \affil[1]{ QuantScape Inc. QuantScape Inc., 4-11-18, Manshon-Shimizudai, Meguro, Tokyo, 153-0064, Japan }
          
\affil[2]{ The School of Electrical Engineering and Informatics, Institut Teknologi Bandung (STEI-ITB), Jl. Ganesha No.10, Bandung, Indonesia }
     
\affil[3]{ Research Collaboration Center for Quantum Technology 2.0, BRIN-ITB-TelU, Indonesia }
     
\affil[4]{ ITB Research Center on ICT (PPTIK-ITB) }

\affil[*]{ hikaruwakaura@gmail.com }

\affil[/]{rahmat.mulyawan@itb.ac.id}  
  
\affil[+]{suksmono@itb.ac.id}

\keywords{ Quantum computer, machine learning, Kolmogorov-Arnold Network }

\begin{abstract}

Quantum imaginary time evolution (QITE) is a powerful method to derive the ground states of the systems.  
Only the damping of quantum states leads it; hence, reaching the ground state is guaranteed by nature without any external manipulation. 
Numerous QITE methods by many groups are used to improve speed and accuracy, derive excited states, and solve combined optimization problems. 
However, the QITE methods have not been used for quantum machine learning to predict the ideal values for multiple input values.  
Therefore, we propose a method for applying QITE methods for quantum machine learning and demonstrate fitting problems of elementary functions and classification problems on a 2-D plane. 
As a result, we confirmed that our method was more accurate than a quantum neural network in solving some problems. 
Our method can be used for other quantum machine learning algorithms; hence, it may be the milestone for applying QITE to quantum machine learning.

\end{abstract}

\begin{document}      

\flushbottom 
\maketitle 
%
%
\thispagestyle{empty}

\section{Introduction}\label{1}

In 1982, Richard P. Feynman first introduced the concept of quantum computing, laying the theoretical foundation for quantum computers \cite{feynman_simulating_1982}. Since then, several groundbreaking quantum algorithms have emerged, including Grover's algorithm for database search \cite{2003quant.ph..1079L} and Shor's algorithm for integer factorization \cite{365700}. However, despite these theoretical advancements, practical implementation was hindered for years due to limitations in available quantum hardware. Early quantum systems lacked the robustness and scalability needed for real-world applications, slowing progress in algorithm development. 
This situation began to change around 2015, when IBM launched publicly accessible quantum computers via cloud platforms, significantly accelerating research in the field. Today, superconducting qubits and ion-trap systems represent the leading quantum hardware platforms. At the same time, other promising qubit technologies are under active investigation, such as Majorana fermions \cite{Mezzacapo_2013}, nitrogen-vacancy (NV) centers in diamond lattices \cite{doi:10.1126/science.1139831}, and Rydberg atoms \cite{2020arXiv201210614W}. While these alternatives offer the potential for more compact and scalable quantum computers, they remain at various stages of research and development. 
Alongside hardware advancements, the field of quantum algorithms has flourished since 2015. A breakthrough came with Dr. Alán Aspuru-Guzik's work on Variational Quantum Algorithms (VQAs) \cite{Kassal2011}, which are well-suited for Noisy Intermediate-Scale Quantum (NISQ) devices. This led to the creation of several influential algorithms, including the Variational Quantum Eigensolver (VQE) \cite{McClean_2016}, Adaptive VQE \cite{2019NatCo..10.3007G}, and Multiscale Contracted VQE (MCVQE)  \cite{2019arXiv190608728P}, as well as a variety of Variational Quantum Machine Learning algorithms \cite{2019QS&T....4a4001K, 2019Natur.567..209H, 2021PhRvP..16d4057B, 2022PhRvA.106b2601A, 2020PhRvL.125j0401W, 2022arXiv220211200K, 2022arXiv220608316Y, PhysRevA.98.032309}.
These developments highlight the growing synergy between quantum hardware and algorithm research and the critical need for tools to simulate and optimize quantum systems' time evolution.

In this context, the quantum imaginary time evolution (QITE) method is proposed \cite{Motta_2019,mcardle_variational_2019}. 
This method and its family are proposed to derive the exact ground energy using the damping of quantum systems by noises.  
The primary method of QITE is Variational QITE (VarQITE), which can be separated into two types; one propagates the systems by solving nonlinear functions for parameters \cite{Motta_2019,mcardle_variational_2019} and another one optimizes parameters to emulate imaginary time propagators \cite{Ledinauskas_2023,Amaro_2022} called.  
Recently, a nonvariational method like the Grover algorithm called probabilistic imaginary time evolution \cite{2022PhRvR...4c3121K} is demonstrated.   
QITE is applied for machine learning on the quantum neural network using a neural network state and some techniques \cite{2025arXiv250322570K,Ledinauskas_2023}.  
          
However, the practical application of QITE on multiple data of even toy models has not been reported.           
Therefore, we present a method for applying QITE to quantum machine learning algorithms, using the Adaptive Variational Quantum Kolmogorov-Arnold Network as an example.      
Adaptive VQKAN \cite{2025arXiv250321336W} is one of 
 VQKAN \cite{Wakaura_VQKAN_2024}, and  VQKAN is one of the variational ways of KAN \cite{2024arXiv240419756L} which is a novel layer network to emulate the evolution of neurons and synapses.       
We chose this method because it has a small number of parameters.     
As a result, the loss functions decrease, and prediction accuracy improves as the  systems propagate by imaginary time.      
Besides, the accuracy of the prediction is better than that of the quantum neural  network for some problems.    
This result is applicable to other methods of quantum machine learning, such as other VQKAN families \cite{2025arXiv250322604W} and quantum neural networks.   
   
Section \ref{1} is the introduction, section \ref{2} describes the method, section \ref{3} describes the result of fitting problems of elementary functions and classification problems, and section \ref{7} is the concluding remark.

\section{Method}\label{2}

In this section, we describe the optimization method for VQKAN using Variational Quantum Imaginary Time Evolution (VarQITE). 
First, we describe the details of VQKAN. 
VQKAN is the quantum version of VQKAN that emulates synaptic learning as a layer of ansatz of Variational Quantum Algorithms.  
We use Adaptive VQKAN because it has a small number of parameters unless the ansatz has over 8 adaptive operators. 
Adaptive VQKAN is VQKAN uses adaptive ansatz for the input vector at layer n for $ m $-th input data $ _n{\bf x} ^m $. 
 
\begin{equation}   
{\Phi}_n^A = M \prod_{p = 0 }^{N _p ^n -1 } exp (i P_p (j, k) \phi_{j k}^{n p} (_n{\bf x} ^m)).     \label{Phia}  
 \end{equation}     
 
whose $ P_p (j, k) $ is the p-th operator on j and k-th qubits, and $ N_p ^n $ is the number of operators on the n-th layer, respectively, for parameter function  
 
\begin{equation}    
\phi_{j k}^{n p} (_n{\bf x} ^m) = \sum_{i \in \{0, dim (_n{\bf x} ^m) \}}^{N_d^n -1 } 2 acos (E_f (_n x_i ^m)  +\sum_{s = 0 }^{N_g  
-1 }\sum_{l = 0 }^{N_s -1 } c_s^{n p j k} B_l (_n x_i ^m)) 
\end{equation}

whose $ c_s^{n p j k} $ and $ B_l (_n x_i ^m) $ are the parameters to be trained, initialized into 0 and spline functions at layer n whose domains are $ [  - 1 , 1  ] $, respectively, the same as classical KAN.  
Then, $ N_d^n $ is the number of inputs for layer $ n $, $ N_g $ is the number of grids for each gate, and $ N_s $ is the number of splines, respectively.
$ E_f (_n x_i ^m) = _n x_i ^m / (exp (-_n x_i ^m) + 1) $ is the  Fermi-Dirac expectation energy-like value of the distribution.      
The component of $ _n{\bf x} $ is the expectation value of the given  observable for the calculated states of qubits.
initial state $ \mid \Psi_{ini} (_1{\bf x} ^m) \rangle $ is $ \prod _{ j = 0 } ^{ N _q-1 } Ry^j (acos(2 _1{\bf x}_j ^m - 1) + 0.5 \pi ) \mid 0 \rangle ^{ \otimes N _q } $ for each input $ m $. 
$ Ry^{j} (\theta) $ is $ \theta $ degrees angle rotation gate for y-axis on qubit $ j $.  
The final state of Adaptive VQKAN is,   
      
 \begin{equation} 
\mid \Psi (_1{\bf x} ^m) \rangle = \prod_{n = 1}^{num. ~ of ~ layers ~ N_l}{\Phi}_n^{A}M \mid \Psi_{ini} (_1{\bf x} ^m) \rangle. 
 \end{equation}  
    
The result is readout as a form of the Hamiltonian expectation value $ H $, and the loss function is calculated as follows,
\begin{eqnarray} 
l_m &=& | \langle \Psi (_1{\bf x} ^m) | H | \Psi (_1{\bf x} ^m) \rangle-f^{aim} (_1{\bf x} ^m) | \\\label{loss} \nonumber
 L &=& \sum_{m = 0}^{num. ~ of ~ samples ~ N -1 } a_m l_m \\\nonumber 
\end{eqnarray}      
    
where $ f^{aim} (_1{\bf x} ^m) $ is the aimed value  of sampled point m and $ l_m $ is the loss function (absolute distance) of point $ m $,  respectively. Hamiltonian takes the form $H = \sum_{j = 0}^{N_o -1 }\theta_j P_j $ for the product of the Pauli matrix $P_j$, consisting of the Pauli matrix $X_j, Y_j, Z_j$. $ N_o $ is the number of $P_j$ in Hamiltonian.     
  
The procedure to choose the new term is two ways: 1.     choose the term from the term pool that has the most significant absolute value of gradient in case the term is included at the end of ansatz 2.     choose the term from the term pool that have the most minor loss function in case the term is included at the end of ansatz  \cite{2024npjQI..10...18D}.     
The parameter for the new term is 0 for all $ c_s^{n p j k} $ s, and no new term is added in case the value of the loss function is decreased by adding no terms in the term pool for case 2.

We use VarQITE to optimize parameters.    
VarQITE can be applied to deriving energy levels of quantum systems and quantum machine learning, using some techniques. 
 
VarQITE is the method to propagate the quantum systems by optimizing parameters for damping operators $ e ^{ - \delta \tau H } $ where $ \delta \tau $ is unit imaginary time and $ H $ is the Hamiltonian, respectively.  
In detail, VarQITE optimizes the following equation for each time step and the state. 
 
\begin{equation} 
l ^{ VarQITE } _ m = \mid 1 - \frac { \langle \Psi _{ \tau }  (_1{\bf x} ^m) | e ^{ - \delta \tau H  } | \Psi  _{ \tau - 1  } (_1{\bf x} ^m) \rangle }{ \sqrt{ cosh2 \delta \tau - sin h 2 \delta \tau \langle \Psi _{ \tau  - 1 }  (_1{\bf x} ^m) |   H  | \Psi  _{ \tau - 1  } (_1{\bf x} ^m) \rangle } }  \mid . 
\label{ q i t e }\end{equation}           
              
The quantum computers can not process the imaginary propagators; hence, they must be decomposed into the Taylor series as follows,        
                           
\begin{equation}          
e ^{ - \delta \tau H  } = \sum _{ j = 0 } \frac{ 1 }{ j ! } (- \delta \tau H ) ^ j.  
\end{equation}                
In case the Hamiltonian $ H $ is a single Pauli operator, the product of term includes $ H $ can be transformed to,    
    
\begin{equation} 
e ^{ - \delta \tau H  } = cos h \delta \tau - H sin h  \delta \tau   
\end{equation}   
    
hence, eq. (\ref{ q i t e }) can be calculated using the swap test and the Hadamard test.   
We minimize the whole loss function     
 
\begin{eqnarray}    
 L ^{ VarQITE } &=& \sum_{m = 0}^{num. ~ of ~ samples ~ N -1 } a_m l_m ^{ VarQITE } \\\nonumber    
\end{eqnarray}       

 for the coefficient $ a _ m = (1 - f^{aim} (_1{\bf x} ^m) + 1 / N) / (2 + 1 / N) $ the same as that of eq. \ref{loss}.    
We assume $ \delta \tau $ is 0.1 and propagate the system growing ansatz per 2.0 $ \tau $ and increasing the number of grids $ N_s $ 4 per 0.1 $ \tau $, respectively.  
We use a swap test to calculate the above loss function.  
Optimization of parameters for first step is initiated from 0 for all components and random values from 0 to 0.1 are added before each step of optimization of parameters on VarQITE. 
If the loss function becomes $ 10 ^{-16} $ or below, the Adaptive VQKAN process converges. 
We assume $ N = 10, N_l = 1, N_q = 4, N_g = 8 $ and $ N_s = 8 $ on initial state, respectively.  
We use blueqat SDK \cite{Kato} for numerical simulation of quantum calculations and COBYLA of scipy to optimize parameters but to declare the use of others. 
We assume that the number of shots is infinite.    
All calculations are performed in Jupyter notebook with Anaconda 3.9.12 and Intel Core i7-9750H.
 Evaluation method of each result is hold-out method.   

\section{ Result }\label{3}  
   
 In this section, we describe our result of learning by VarQITE for fitting and classification problems. 
The operator pool consists of all combinations of one body Pauli matrices X, Y, Z and two body Pauli matrices $ XX, XY, XZ, YY, Y Z, Z Z $ for all combinations of indices without the conjugate coupling.

\subsection{ Fitting }

First, we describe the result of the fitting problem.  
We performed the Adaptive VQKAN on a fitting problem of the following equation, exponential function $ exp ( (x _ 1 - x _ 2) ^ 2  / 2 x _ 0) $, logarithmic function $ log (x _ 0 / x _ 1) $, fractional function $ 1 / (1 + x _ 0 x _ 1) $, radius of sphere which center is zero point on 10 sampled points and predicted the values of 50 test points.    
The target function is defined as:  
    
\begin{equation}    
 f^{\rm aim} ({\bf x}) = \exp\left(\sin(x_0^2 + x_1^2) + \sin(x_2^2 + x_3^2) \right). \label{last} 
 \end{equation}  
  
 Here, $ x _ i = \sqrt{ 1 - (2 _ 1{\bf x} ^m _i - 1) ^ 2 } $ for $ i = 0, 1, 2, 3 $.
  
\begin{figure}                
 
\includegraphics[scale= 0.25    ]{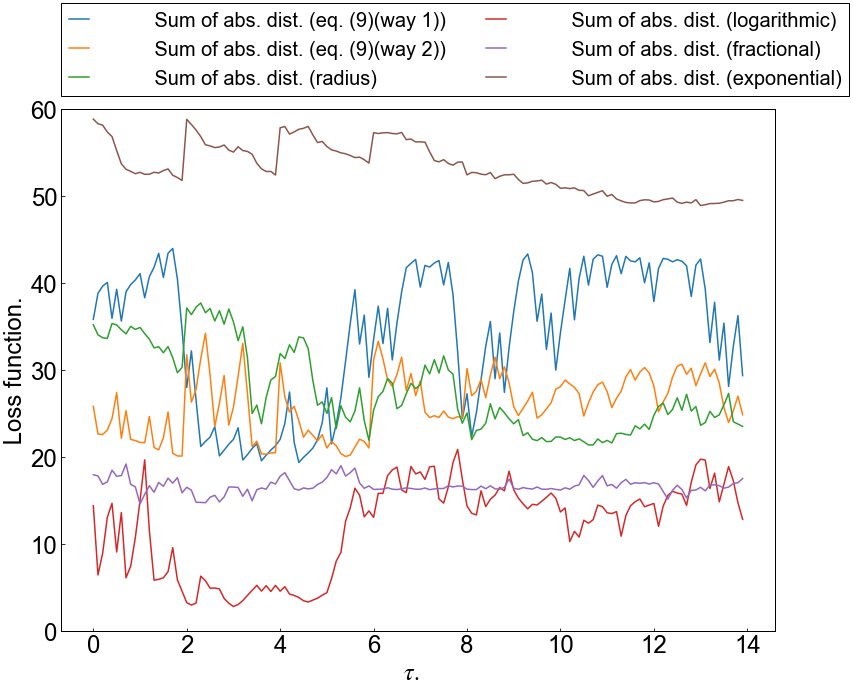}    
 
 \caption{ Imaginary time $ \tau $ vs. sum of absolute distances (loss function of test points) on the fitting of eq.  (\ref{last}), the radius of the sphere, logarithmic function, fractional function, and exponential function. } \label{ i t l }  
       
\end{figure}

\begin{figure}               
 
\includegraphics[scale= 0.25    ]{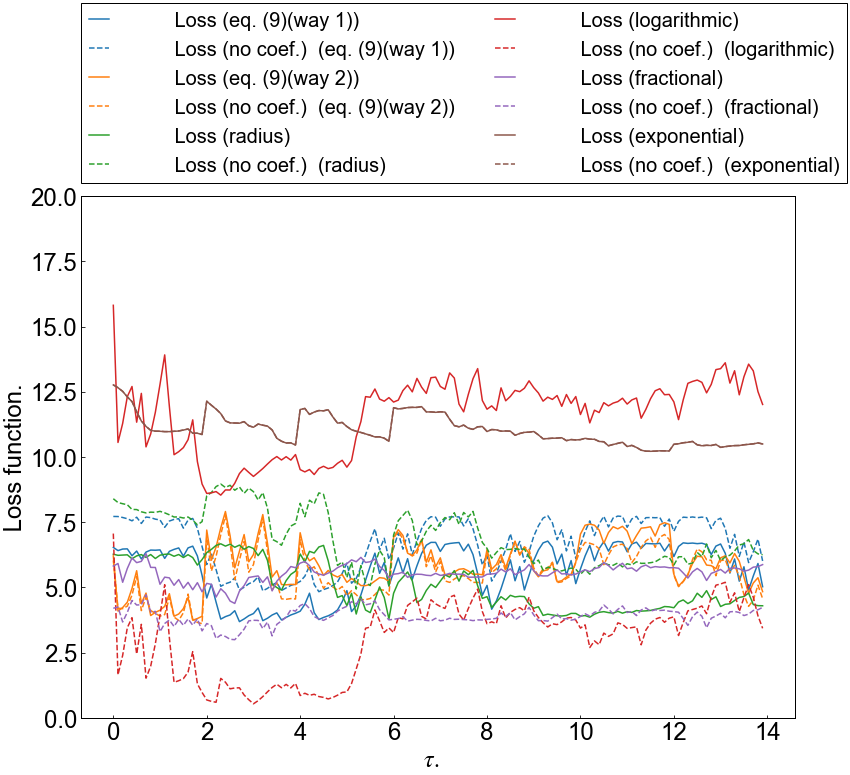}   

 \caption{ Imaginary time $ \tau $ vs. loss function and loss function without coefficie nts on the fitting of eq.  (\ref{last}), the radius of the sphere, logarithmic function, fractional function, and exponential function. } \label{ i t l d }     
      
\end{figure}

 $ _n{\bf x} ^m _i = 0.5 (\langle \tilde{ \Psi }    (_1{\bf x} ^m) | Z_i | \tilde{ \Psi }  (_1{\bf x} ^m) \rangle + 1)$ for the state calculated by n-th layer $ | \tilde{ \Psi }  (_1{\bf x} ^m) \rangle $, with $N_d^n = 4$ and $\dim(_n{\bf x} ^m) = 4$ for all layers and calculations, and the Hamiltonian is $Z_0 Z_1 + Z_2 Z_3$. 
The range of $ x _ i $ is $ [  - 1 , 1  ] $ for eq. (\ref{last})  and radius of sphere, and $ [  0, 1  ] $ for others.       
The number of layers $ N_l = 1 $ and initial ansatz is $ X_ 1 $,  respectively.          
We show the loss function and the sum of absolute distances for imaginary time of eq.  (\ref{last}), exponential function $ exp ( (x _ 1 - x _ 2) ^ 2  / 2 x _ 0) $, logarithmic function $ log (x _ 0 / x _ 1) $, fractional function $ 1 / (1 + x _ 0 x _ 1) $, radius of sphere which center is zero point in Fig. \ref{ i t l d } and Fig. \ref{ i t l }, respectively. 
 
\begin{figure}                
  
\includegraphics[scale= 0.25    ]{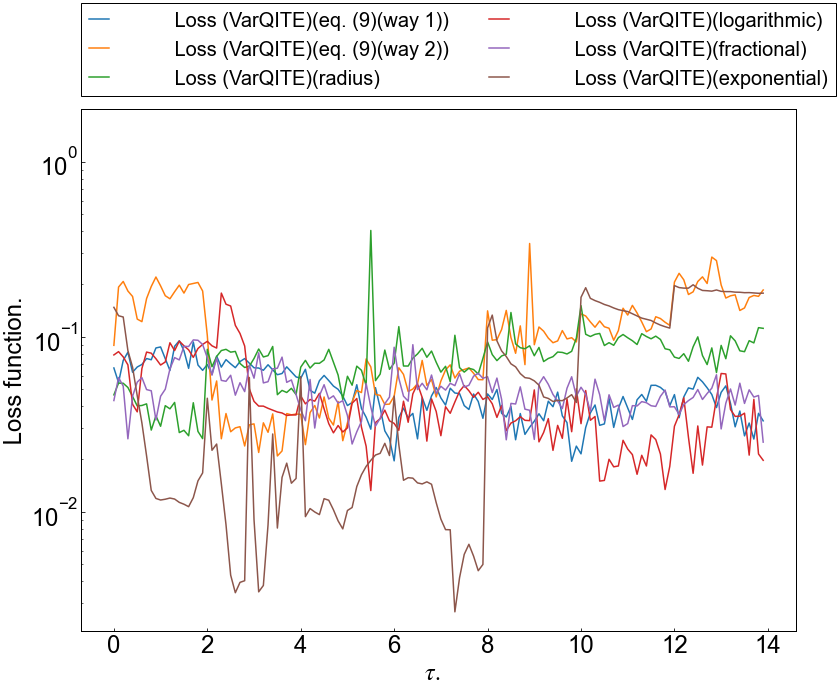}  

 \caption{ Imaginary time $ \tau $ vs. $  L ^{ VarQITE } $ on fitting of eq.  (\ref{last}), the radius of the sphere, logarithmic function, fractional function, and exponential function. } \label{ i t l p }   
       
\end{figure}

\begin{figure}                   
  
\includegraphics[scale= 0.25    ]{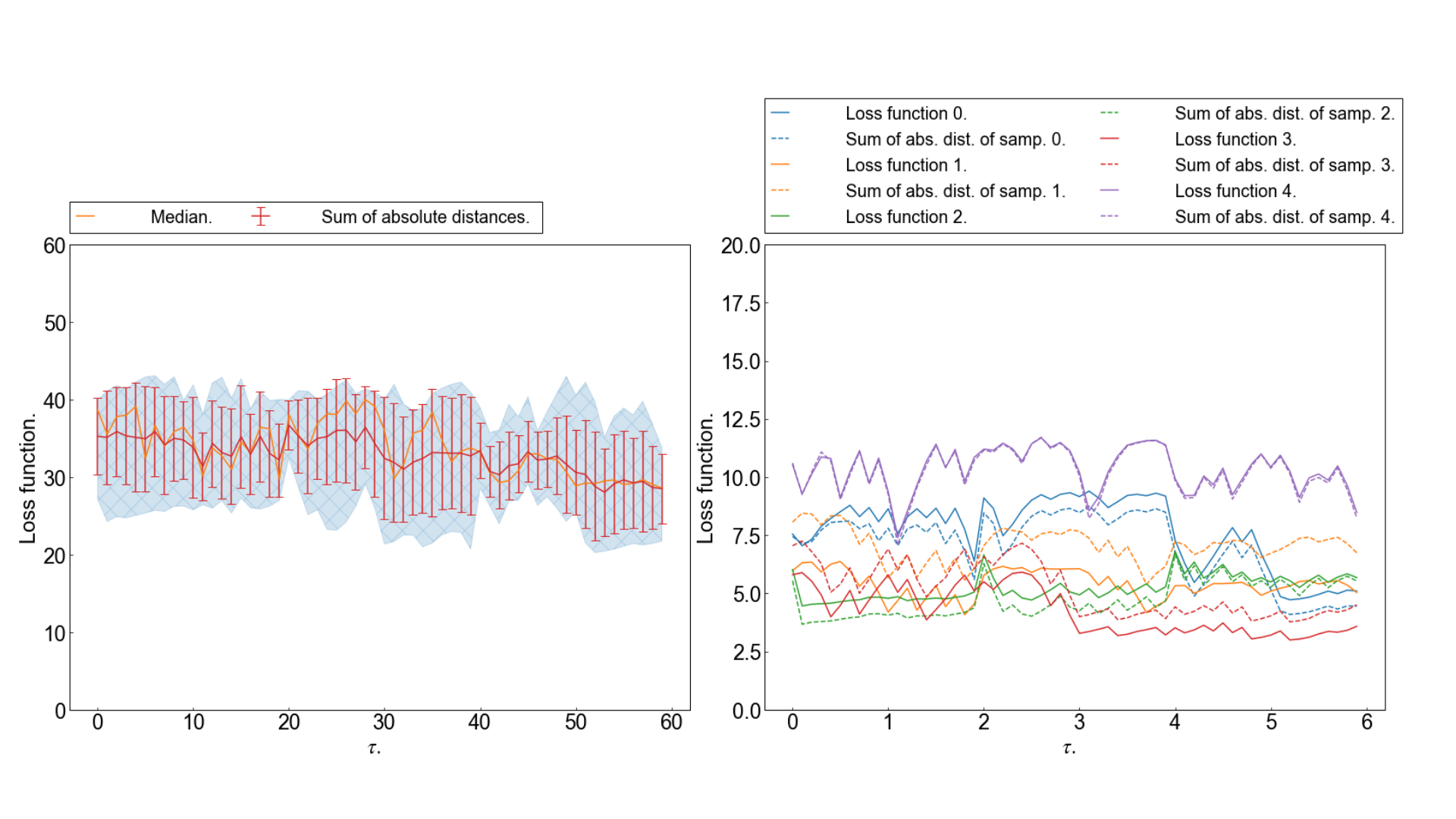}     
 
 \caption{ (Left) Imaginary time $ \tau $ vs. the average and median for 5 attempts of the sum of absolute distances on the fitting of eq.  (\ref{last}). The error bar indicates the standard deviation. (Right) Imaginary time $ \tau $ vs. loss function and loss function without coefficients for 5 attempts of the sum of absolute distances on the fitting of eq.  (\ref{last}). } \label{ i t l i t }     
          
\end{figure}

The values of loss function decrease gradually dwelling, and the sum of absolute distances decrease gradually dwelling for the domain $ \tau $ is below 6, respectively.   
The absolute distances of eq. (\ref{last}) and radius decrease slower than others.  
The more input values there are, the slower the convergence is.      
 Besides, the sum of absolute distances rebounded, and convergence speed became slower in domain  $ \tau > 6 $.    
This is because VarQITE's loss functions need a larger number of trials to derive parameters that emulate imaginary time propagators while maintaining accuracy.    
 The loss functions of VarQITE rise gradually from there except logarithmic function and fractional function, as shown in Fig. \ref{ i t l p }.    
 According to the previous paper \cite{2025arXiv250322604W}, the minimum values of sum of absolute distances of  eq. (\ref{last}) logarithmic function and fractional function are smaller than average of Quantum Neural Network.     
The minimum of the fractional function is smaller than that of EVQKAN.   
The result for demonstration fitting on eq. (\ref{last}) 5 times for each sampled point, and common test points also showed slow convergence, as shown in Fig. \ref{ i t l i t }. Both the sum of absolute distances and loss function decrease gradually on average.    
Even though the few number of terms in adaptive ansatz, the damping of the systems leads to the optimization of the parameters as the networks of synapses.

\subsection{ Classification  }
Next, we describe the result of the classification problem.   
We present the results of solving the classification problem for points on a 2-D plane. A point is assigned a label of $+1$ if it is above the function $f$ and $-1$ if it is below $f$. The  function $f$ is defined as:  
 \begin{equation}  
f(x) = \exp(d_0 x_0 + d_1) + d_2 \sqrt{1-d_3 x_0^2} + \cos(d_4 x_0 + d_5) + \sin(d_6 x_0 + d_7) 
 \end{equation} 
where $d_k$ represents random coefficients between 0 and 1 for the various cases. The loss function for the classification is given as follows: 
    
\begin{equation}   
f^{\rm aim} =  
\begin{cases}     
-1 & \text{if} f \geq x_1 \\   
1 & \text{if} f < x_1 \\  
\end{cases}
 \end{equation}  
 
\begin{figure}               
    
\includegraphics[scale= 0.25    ]{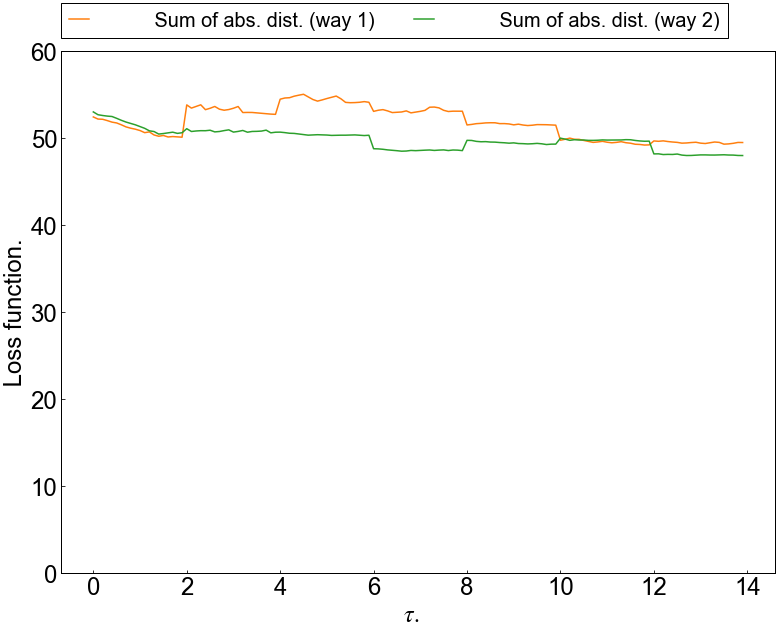}  
  
 \caption{ Imaginary time $ \tau $ vs. sum of absolute distances (loss function of test points) on a classification problem. } \label{ i t l c }  
         
\end{figure}

\begin{figure}                 
 
\includegraphics[scale= 0.25    ]{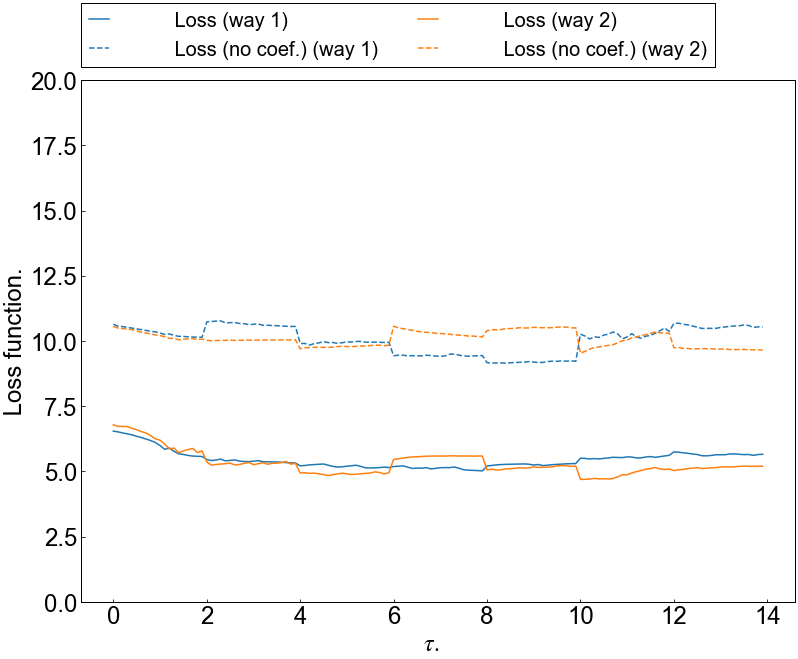}    

 \caption{ Imaginary time $ \tau $ vs. loss function and loss function without coefficients on a classification problem. } \label{ i t l c d }       
      
\end{figure}

The range of $ x _ i = \sqrt{ 1 - (2 _ 1{\bf x} ^m _i - 1) ^ 2 } $ is $ [  - 1, 1  ] $ and $ _n{\bf x} ^m$ is $ [  0, 1  ] $, respectively. 
We show the results of the classification for different cases: using Adaptive VQKAN with $ \dim(_n{\bf x} ^m)  = 2 $ for $n > 2$ for a single attempt in case the range of $ \tau $ is 0 to 14 for way 1 and 2, respectively.     
 We show the loss function and the sum of absolute distances for imaginary time in Fig. \ref{ i t l c d } and Fig. \ref{ i t l c }, respectively.
The sum of absolute distances decreases gradually in both ways.  
Adaptive ansatz grows appending the $ Y _ 1 $ term mainly in both ways, too. 
   
\begin{figure}                  
\includegraphics[scale= 0.25    ]{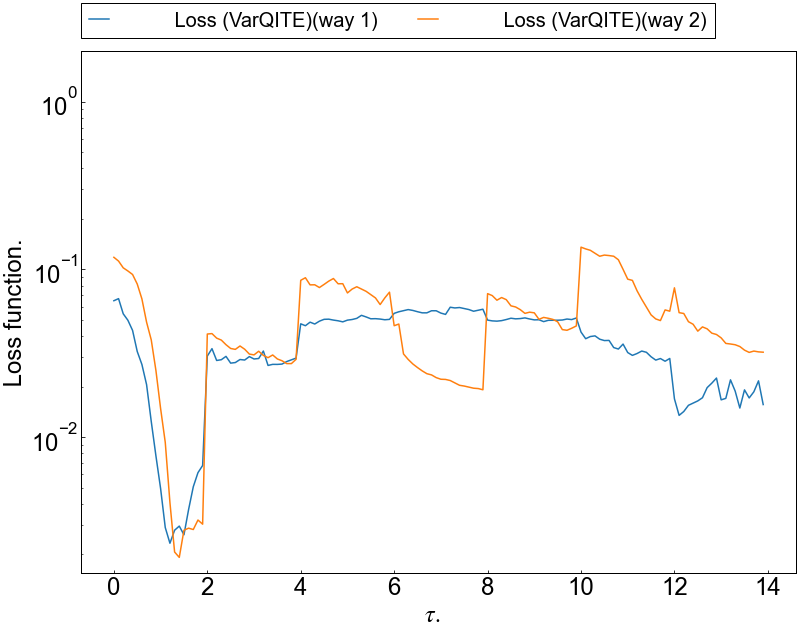}   
  
 \caption{ Imaginary time $ \tau $ vs. $  L ^{ VarQITE } $ on classification problem. } \label{ i t l c p }    
        
\end{figure}

However, their values are larger than those of the values as a result of ordinary Adaptive VQKAN.   
The optimization of parameters could not accurately emulate the imaginary time propagation as shown in Fig. \ref{ i t l c p }.  
 
\begin{figure}               
 
\includegraphics[scale= 0.25    ]{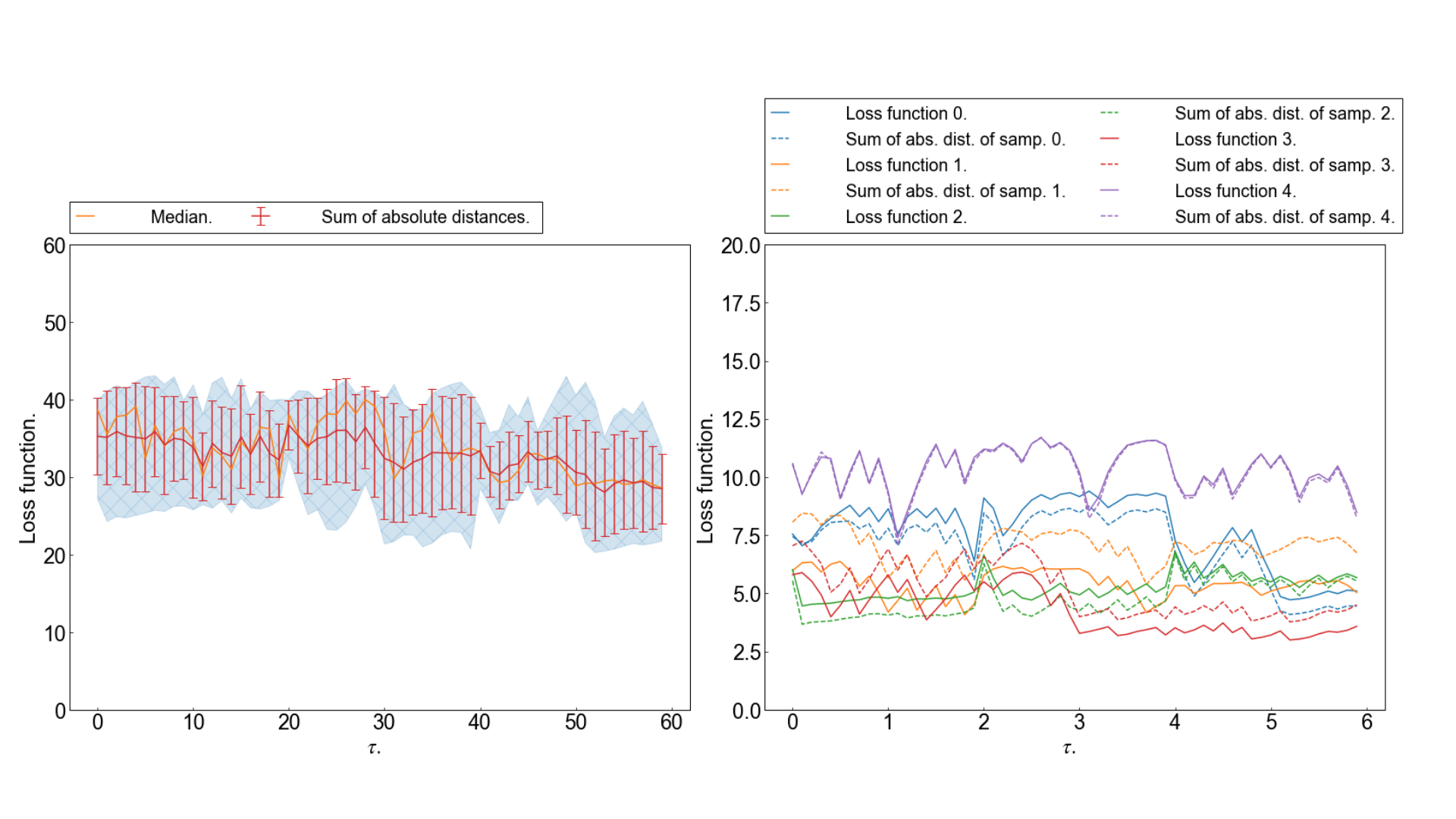}   

 \caption{ (Left) Imaginary time $ \tau $ vs. the average and median for 5 attempts of the sum of absolute distances on a classification problem. (Right) Imaginary time $ \tau $ vs. loss function and loss function without coefficients for 5 attempts of the sum of absolute distances on a classification problem. } \label{ i t l c i t }     
         
\end{figure}

 The result for demonstration of classification 5 times for each sampled points and common test points showed also slow convergence as shown in Fig. \ref{ i t l c i t }. 
 
Adaptive VQKAN is not good at classification even though using damping of the system for optimization. Optimization may reach practical accuracy concerning the absolute distances if we take more imaginary time. 

Sign problem \cite{2022Natur.603..416H} of VarQITE may exist on this Adaptive VQKAN too. 

Imaginary time evolution damps the systems faster on ordinary VarQITE for root optimization \cite{Amaro_2022}, Multiple Knapsack Problem \cite{2025arXiv250412607L} and Probabilistic imaginary-time evolution for deriving energy levels \cite{2022PhRvR...4c3121K} and solving advection-diffusion equation \cite{2024arXiv240918559H}.      
 
Neither VarQITE nor Probabilistic imaginary-time evolution uses a subspace search method; hence, the subspace search method may lower the speed of convergence and accuracy of emulating imaginary time propagators.                
Besides, the rebound of the sum of absolute distances by growing ansatz may be due to the initial parameters of added terms \cite{Amaro_2022}.

\section{ Concluding remarks }\label{7}

In this paper, we reveal that VarQITE can optimize Adaptive VQKAN. This means that quantum machine learning can be performed by quantum imaginary time evolution without special techniques for some methods.    
However, the slow damping of systems and rebound of the absolute distances remained problems.     
Optimizing Adaptive VQKAN by different methods, such as Probabilistic imaginary-time evolution, and comparing the results are the next problems.  
Making initial parameters using other methods to avoid rebound is also important.

\bibliography{mainwakaura2}


 \end{document}